\documentclass[referee]{raa} 
\usepackage{graphicx,times}
\usepackage{natbib}
\usepackage{amssymb,amsmath}
\bibpunct{(}{)}{;}{a}{}{,}
 
\usepackage[a4paper=true,dvipdfm=true,pagebackref=true]{hyperref}
\hypersetup{pdftitle = The title of my PDF, pdfauthor = My name, pdfsubject= The subject, pdfkeywords = keyword1 keyword2 keyword3} 
\hypersetup{colorlinks = true, linkcolor = green, anchorcolor = red, citecolor = blue, filecolor = red, pagecolor = red, urlcolor = red}

\begin{document}

\title{LAMOST Medium-Resolution Spectral Survey of Galactic Nebulae
(LAMOST-MRS-N): Subtraction of Geocoronal H$\alpha$ Emission}

   \volnopage{Vol.0 (20xx) No.0, 000--000}      
   \setcounter{page}{1}          
 
   \author{Wei Zhang
      \inst{1,2}
   \and Hong Wu
      \inst{1,2}
   \and Chaojian Wu
      \inst{1,2}
   \and Juanjuan Ren
      \inst{3,2}
   \and Jianjun Chen
      \inst{1,2}
   \and Chih-Hao Hsia
      \inst{4}
   \and Yuzhong Wu
      \inst{1,2}
   \and Hui Zhu
      \inst{2}
   \and Jianrong Shi
      \inst{1,2}
   \and Zhongrui Bai
      \inst{1,2}
   \and Zhaoxiang Qi
      \inst{5}
   \and Yongheng Zhao
      \inst{1,2}
   \and Yonghui Hou
      \inst{6,7}
   }
 
   \institute{CAS Key Laboratory of Optical Astronomy, National Astronomical Observatories,  Chinese Academy of Sciences, Beijing 100101, China; {\it xtwfn@bao.ac.cn}\\
        \and
             National Astronomical Observatories, Chinese Academy of Sciences, Beijing 100101, China\\
        \and
             CAS Key Laboratory of Space Astronomy and Technology, National Astronomical Observatories, Chinese Academy of Sciences, Beijing 100101, China\\
        \and
             State Key Laboratory of Lunar and Planetary Sciences, Macau University of Science and Technology, Taipa, Macau, China\\
        \and
             Shanghai Astronomical Observatory, Chinese Academy of Sciences, 80 Nandan Road, 200030 Shanghai 200030, China\\
        \and
             Nanjing Institute of Astronomical Optics \& Technology, National Astronomical Observatories, Chinese Academy of Sciences, Nanjing 210042, China\\
        \and
             School of Astronomy and Space Science, University of Chinese Academy of Sciences, Beijing 100049, China\\
\vs\no
   {\small Received~~20xx month day; accepted~~20xx~~month day}}
 
        
\label{firstpage}

\abstract{
We introduce a method of subtracting geocoronal H$\alpha$ emissions from the
spectra of LAMOST medium-resolution spectral survey of Galactic nebulae (LAMOST-MRS-N).
The flux ratios of the H$\alpha$ sky line to the adjacent OH $\lambda$6554
single line do not show a pattern or gradient distribution in a plate.  More
interestingly, the ratio is well correlated to solar altitude, which is the
angle of the sun relative to the Earth's horizon. It is found that the ratio
decreases from 0.8 to 0.2 with the decreasing solar altitude from -17 to -73
degree. Based on this relation, which is described by a linear
function, we can construct the H$\alpha$ sky component and subtract it from the
science spectrum. This method has been applied to the LAMOST-MRS-N data, and
the contamination level of the H$\alpha$ sky to nebula is reduced from 40\% to
less than 10\%. The new generated spectra will significantly improve the
accuracy of the classifications and the measurements of physical parameters of
Galactic nebulae.
}

   \authorrunning{Wei Zhang, Hong Wu, Chaojian Wu, et al.}            
   \titlerunning{Subtraction of Geocoronal H$\alpha$ Emission}  
 
   \maketitle

\section{Introduction}

Sky subtraction is necessary for the ground-based spectroscopic measurements,
especially for faint objects. For the long-slit spectrographs, the sky can be
derived by interpolating adjacent blank sky regions and subtracted from the
target \citep{Soto-2016}. On the other hand, this method dose not work well for
the multi-fiber spectrographs, but considerable progress has been made in the
last three decades to address sky subtractions, such as beam-switching
\citep{Barden-1993,Puech-2014,Rodrigues-2012}, nod-and-shuffle
\citep{Glazebrook-2001,Sharp-2010}, and principal component analysis
\citep[PCA,][]{Wild-2005,Soto-2016}. As an example, for the Large Area
Multi-Object fiber Spectroscopic Telescope (LAMOST) survey \citep{Wang-96,
Su-Cui-04, Cui-12,Zhao-12,Luo-15}, a master sky spectrum was first constructed
from sky fibers and subtracted from each science spectrum, then the residual
sky in OH bands was further removed using the PCA sky-subtraction method
\citep{Bai-2017}.

LAMOST has started the medium-resolution ($R\sim7500$) survey (LAMOST-MRS) on
October 2018. The blue segment spectrum covers the spectral range
4950--5350\,\AA, while the red one covers 6300--6800\,\AA. A number of
sub-projects are underway simultaneously to achieve the scientific
goals including binarity/multiplicity, stellar pulsation, star formation,
emission nebulae, Galactic archaeology, host stars of exoplanets, open
clusters, and so on \citep{Liu-2020}. As one sub-project among them, LAMOST
medium-resolution survey of Galactic nebulae (LAMOST-MRS-N) focuses on emission
nebulae on the Galactic plane spanning the longitude range $40\degr < l <
215\degr$ and the latitude range $|b| < 5\degr$ \citep{Wu-2020b}.

Five prominent emission lines, including H$\alpha$, [N{\,\sc
ii}]$\lambda\lambda$6548,6584 and [S{\,\sc ii}]$\lambda\lambda$6717,6731, are
in the coverage of the red segment spectrum of the LAMOST-MRS-N.  The spectrum
is rich in informations of oxygen abundances, electron densities, radial
velocities, and velocity dispersions of the interstellar medium (ISM). Besides,
the line ratios are usually used to classify the ISMs into H{\,\sc ii} regions,
supernova remnants (SNRs), and planetary nebulae (PNe) \citep{Sabbadin-1977,
Riesgo-2006}. However, the H$\alpha$ nebular line (H$\rm \alpha_{neb}$) is
blended with geocoronal H$\alpha$ sky line (H$\rm \alpha_{sky}$) which is
the result of solar Lyman $\beta$ scattering by atomic hydrogen in the
Earth's upper atmosphere \citep{Mierkiewicz-2006, Gardner-2017}. The sky
component should be subtracted before we derive the parameters based on the
H$\alpha$ nebular line. Unfortunately, it is impossible to construct the master
spectrum from the sky fibers in the same field or from the adjacent regions, and
to subtract the sky from the target, because the sky fibers cannot be dedicated
due to the fact that there are always diffuse Galactic lights being fed into
the fibers.  Therefore, a novel method of sky subtraction for LAMOST-MRS-N is
needed.

In this paper, we develop a new method of subtracting H$\rm \alpha_{sky}$ from
the science spectrum with the help of the single sky line OH $\lambda$6554.
This method is convenient to be applied to the LAMOST-MRS-N data. In the
following section, we first describe the properties of the sky lines observed
at a dark night (Section 2.1), then we decompose the sky and nebular components
for some plates in which these two components can be well resolved (Section
2.2), and we find there is a good correlation between the H$\rm
\alpha_{sky}/\lambda$6554 flux ratio and  solar altitude (Section 2.3),
finally we construct the H$\rm \alpha_{sky}$ spectrum for each target and
subtract it from the science spectrum in Section 2.4. We summarise the results
in the last section.

\section{Method}
\subsection{Dark Night Sky Spectrum}

\begin{figure}
\centering
\includegraphics[width=\textwidth]{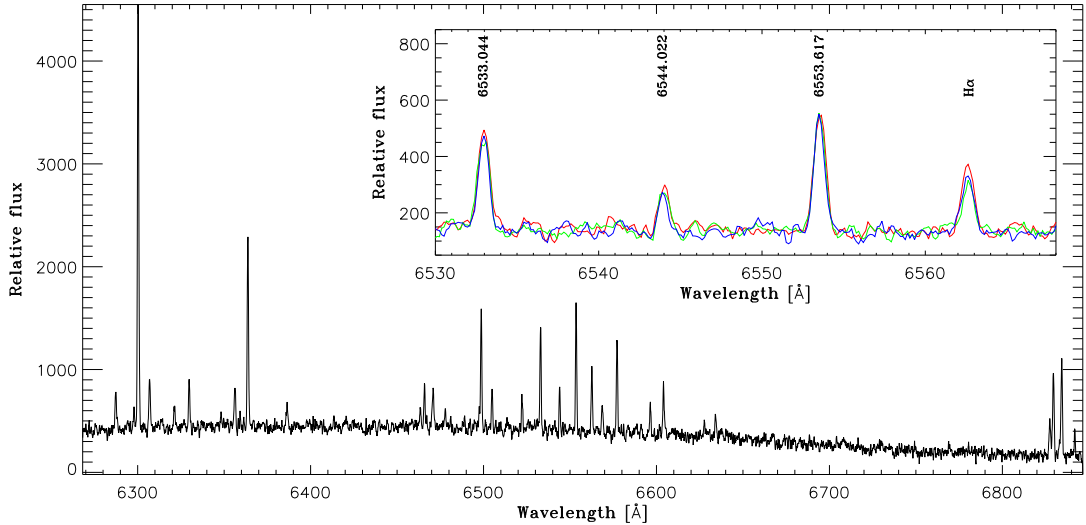}
\caption{An example of median-resolution dark night sky spectrum. The black is the combined spectrum from three exposures. In order to show the details of the lines adjacent to H$\alpha$ in wavelength, we also show the three single exposure spectrum (red, green and blue lines, respectively) in the small window and narrow wavelength coverage.}
\label{fig:skyline_spectra}
\end{figure}

Aiming to explore the details of H$\rm \alpha_{sky}$, such as line intensity,
line dispersion, and the possible correlations with other sky lines, we carried
out a special observation on 2020 November 08, that all fibers are assigned to
blank sky regions instead of stars or galaxies. Besides, this plate has been
set to point to a field at high Galactic altitude (l, b)= (109\fdg2, -30\fdg3)
to avoid the pollution of the diffuse emissions from the Galactic plane. The
field has been observed for a total of 2700s, with the integration time of
900s for each one of the three exposures.

The combined spectrum in the red channel is shown in
Figure~\ref{fig:skyline_spectra}. We can see the strongest line is O{\sc i}
$\lambda$6300. Three single exposure spectra in a narrower wavelength coverage
are shown with different colors in the small window. Besides H$\rm
\alpha_{sky}$, there are three other OH sky lines with the wavelength in air of
6533.044, 6544.022, and 6553.617\,\AA, which are referred to as $\lambda$6533,
$\lambda$6544 and $\lambda$6554 in the following text. We note that
$\lambda$6533, $\lambda$6544, $\lambda$6554 are single lines and have been used
to recalibrate the wavelength for the LAMOST-MRS-N data \citep{Ren-2021}.  The
other reason of analyzing these lines is that they have similar wavelengths to
H$\rm \alpha_{sky}$, therefore the effects of the line spread function,
 wavelength solution, and efficiency along the wavelength can be omitted.

\begin{figure}
\centering
\includegraphics[width=\textwidth]{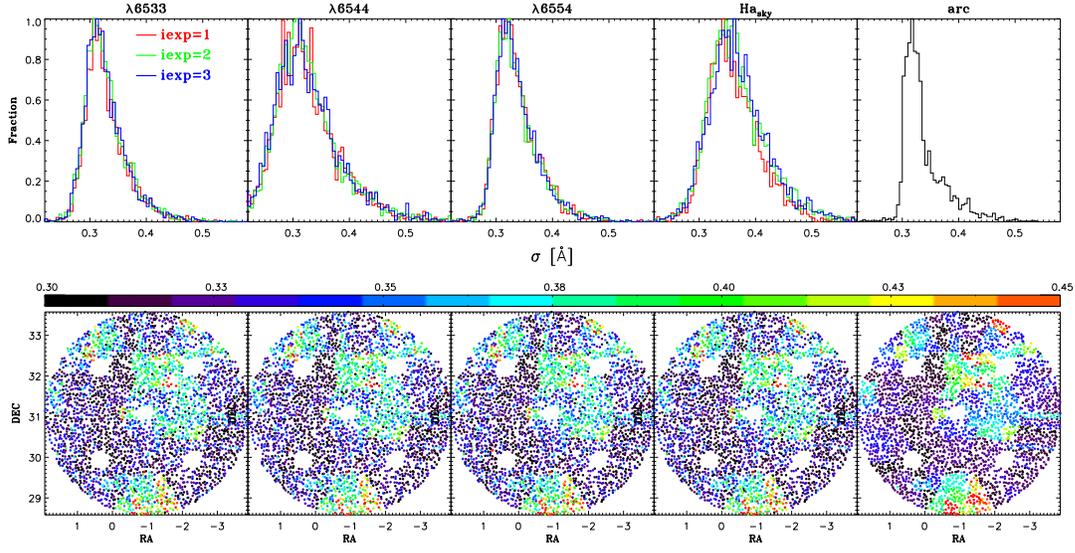}
\caption{Top row: histograms of the line dispersions of four sky lines and that of the mean value of two arc lines. The results of the first, second and third exposure for the sky lines are plotted as red, green and blue lines, respectively.  Bottom row: spatial distributions of the line dispersions.}
\label{fig:skyline_sigma}
\end{figure}

\begin{figure}
\centering
\includegraphics[width=\linewidth]{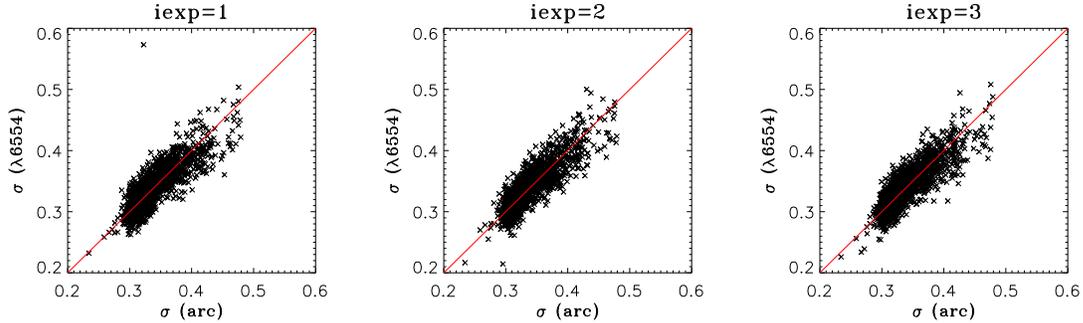}
\caption{Comparison of the line dispersions of arc line to sky line $\lambda6554$. The results of the first, second and third exposure are listed from left to right. The red line is the 1:1 line.}
\label{fig:compare_arc_sky_sigma}
\end{figure}

For each of these four lines, we fit it with a single Gaussian profile and
obtain simultaneously line centroid, line dispersion, and line intensity. In
the upper panels in the left four columns in Figure~\ref{fig:skyline_sigma}, we
checked the histograms of line dispersions of all fibers in the field. The red,
blue and green lines stands for the first, second and third exposure,
respectively. For all the lines, the histograms do not change among three
exposures. However, none of these histogram is a Gaussian profile, instead,
there is an obvious tail at the larger line dispersion. As the sharp arc line
in the Th-Ar lamp spectrum, which is used for the wavelength calibration, can
be assumed to have no intrinsic broadening, the line dispersion is caused by
the instrumental broadening. We show the result of the mean value of two arc
lines ($\lambda$6531 and $\lambda$6584) in the right column, and found the
similar distribution to the sky lines. The tails of the distributions of both
sky and arc lines can be explained according to the spatial distributions shown
in the bottom panels, in which there are some fibers have obvious larger line
dispersions in some spectrographs. The pattern will disappear when the
current problem of the consistency of fiber resolution is resolved. We should
keep in mind that, if we want to study two-dimensional distribution of the
velocity dispersion of the ISM, the instrumental broadening should be removed,
otherwise, the intrinsic broadening from the instrument will affect the result
dramatically.

\begin{figure}
\centering
\includegraphics[width=\linewidth]{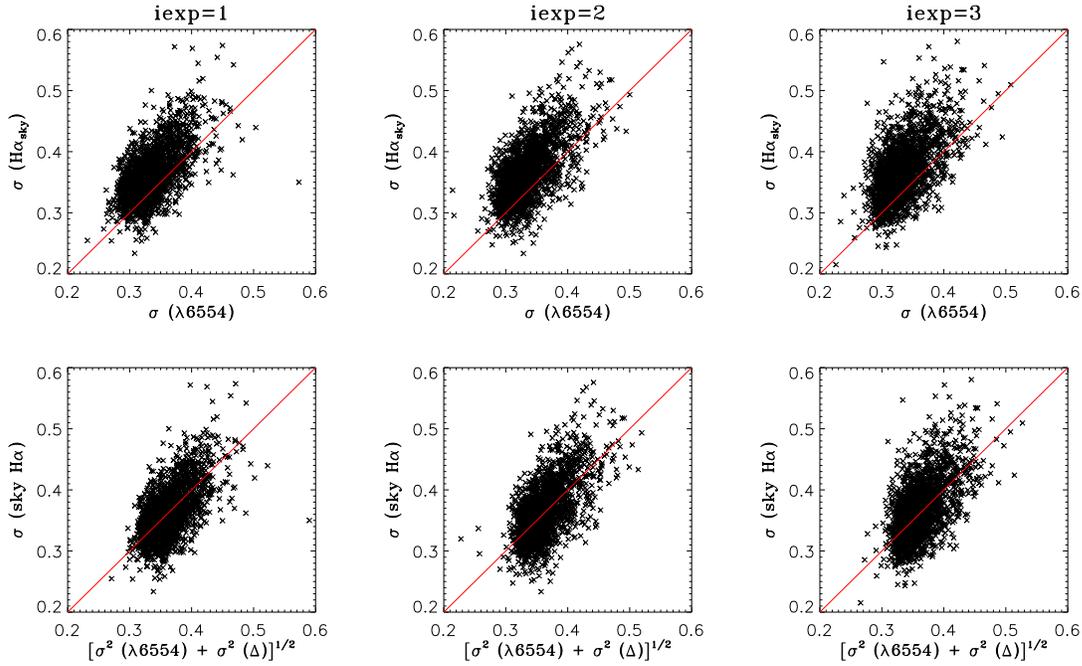}
\caption{Top row: comparison of the line dispersion between $\lambda6554$ and H$\rm \alpha_{sky}$. Bottom row: an extra dispersion of 0.14\,\AA\ has been added to sky $\lambda$6554 to compare to H$\rm \alpha_{sky}$. The red line is the 1:1 line.}
\label{fig:compare_sky_sigma}
\end{figure}

We then compared the line dispersion of the arc line with the sky line
$\lambda$6554 one by one in Figure~\ref{fig:compare_arc_sky_sigma}. It is found
that the line dispersion of $\lambda$6554 is identical with that of the arc
line, indicating the instrumental broadening also dominates the line dispersion
of the sky line $\lambda$6554. However, it is not true for H$\rm \alpha_{sky}$.
As shown in Figure~\ref{fig:compare_sky_sigma}, the line dispersion of H$\rm
\alpha_{sky}$ is systematically higher than that of the sky line $\lambda$6554.
It is found that an extra broadening of 0.14\,\AA\ should be added to the line
dispersion of $\lambda$6554 to match that of H$\rm \alpha_{sky}$. Physically,
geocoronal H$\alpha$ is not a single line, instead, it is composed mainly of
two single lines, with a flux ratio about 2:1 and a wavelength differ of
0.047\,\AA, or equally 2.133 $\rm km\,s^{-1}$ in Doppler shifts
\citep{Mierkiewicz-2006}. With the resolution power of 7500, we cannot resolve
theses two components and a single Gaussian component can fit this line well.
However, the line dispersion is higher than that of a single sky/arc line which
is broadened mainly by the instrumental broadening. This extra broadening
caused by the two components is about 0.02\,\AA, therefore the emission from
the Warm Ionized Medium (WIM) and/or unresolved, faint nebulae might cause the
large difference of the line dispersions between H$\rm \alpha_{sky}$ and the
sky line $\lambda$6554. As revealed by the Wisconsin H$\alpha$ Mapper (WHAM)
survey, this emission is present at some level over the whole sky, even at $|b|
\sim 30^{\circ}$ \citep{Nossal-2001}, where the dark sky plate is targeted.

\begin{figure}
\centering
\includegraphics[width=\textwidth]{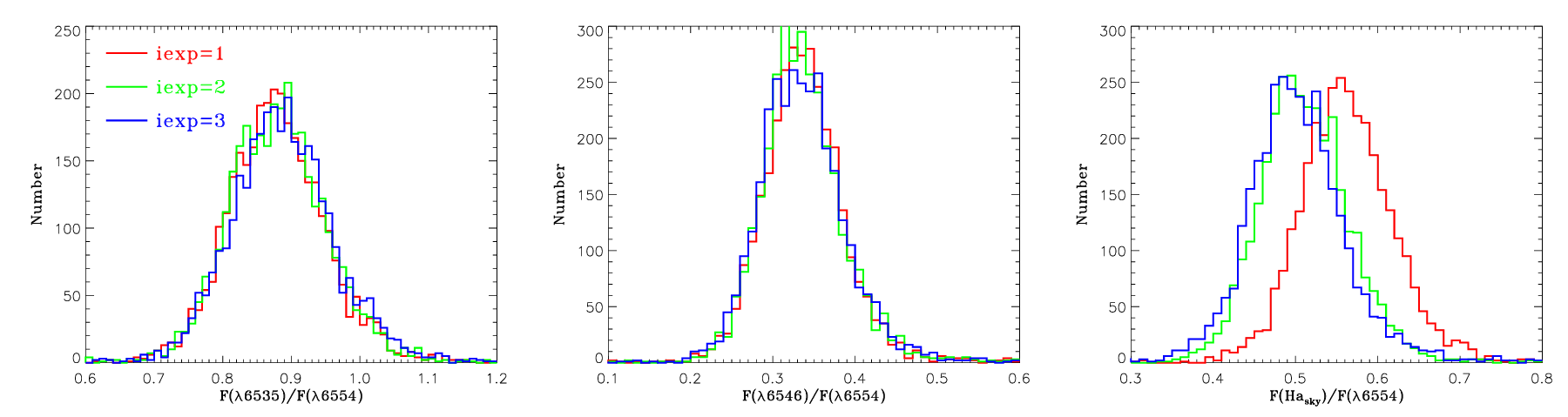}
\includegraphics[width=\textwidth]{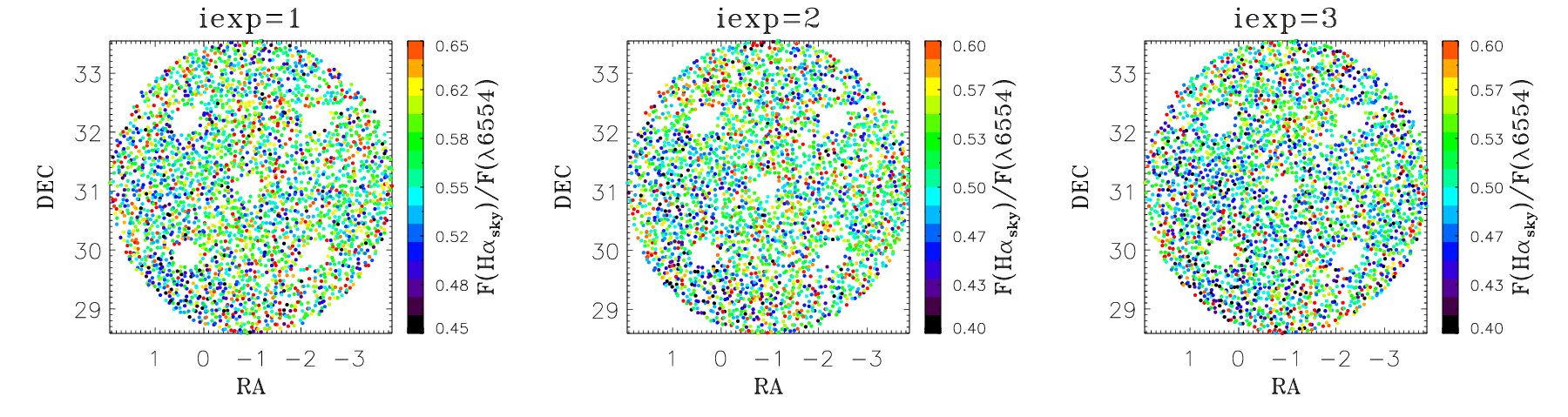}
\caption{Top row: the histograms of $\rm F(H\alpha_{sky})/F(\lambda6554)$ ratio. Colours stands for different exposures. Bottom row: The spatial distribution of $\rm F(H\alpha_{sky})/F(\lambda6554)$ ratio for each single exposure.}
\label{fig:sky_ratio_4lines}
\end{figure}

Finally, we explored whether the F(H$\rm \alpha_{sky}$)/F($\lambda$6554) ratio
changes with time. The histograms of the ratios have been shown in the upper
panels in Figure~\ref{fig:sky_ratio_4lines}. As can be seen, the ratios of
F($\lambda$6535)/F($\lambda$6554) and F($\lambda$6544)/F($\lambda$6554) do not
change with time, while the F(H$\rm \alpha_{sky})$/F($\lambda$6554) ratios from
the second and third exposures change significantly compared to that from the
first exposure. This result indicates that we should first derive the realistic
ratio if we want to use the sky $\lambda$6554 to infer the intensity of H$\rm
\alpha_{sky}$. In the bottom panels, we show the spatial distributions of the
F(H$\rm \alpha_{sky})$/F($\lambda$6554) ratio, and found there is no obvious
pattern or spatial gradient, indicating that a single ratio should be good
enough for a whole plate to describe the H$\rm \alpha_{sky}$ component.
Although the emission from the WIM and/or faint nebulae is blended with the
H$\rm \alpha_{sky}$ component, the line intensity is too low to change the
two-dimensional distribution in the plate (see Section 2.3).

In a summary, the H$\alpha$ sky component has a little large line dispersion
compared to the sky line $\lambda$6554 which is mainly instrumental broadening,
and its line dispersion can be inferred from $\lambda$6554 by adding an
additional broadening of 0.02\,\AA. Besides, the F(H$\rm
\alpha_{sky})$/F($\lambda$6554) ratio can be taken as a constant in a specific
exposure. However, this ratio changes with time, we should calculate the ratio
for each plate.  As the science spectra in one plate are combined from three
exposures, we will calculate one ratio for the plate, instead of three ratios
for these exposures.

\subsection{Science Spectra}

Unlike the dark night sky spectrum, the H$\alpha$ emission line in science
spectrum is composed of sky and nebular components. The straight way is to fit
the H$\alpha$ emission using two Gaussian components. However, it cannot work
well because the two components cannot always be resolved under the resolution
of $R\sim7500$.  Here we define a parameter $\rm RV_{sep}$, which is the radial
velocity separation between the centroids of the two components. This parameter
is not determined from two-component Gaussian fit, instead, it is derived as
$\rm RV_{sep} = |RV([N${\sc ii}$\rm ]\lambda6584)-RV(\lambda6554)|$, based on
the fact that $\rm RV(H\rm \alpha_{sky})\sim RV(\lambda$6554) and $\rm
RV(H\alpha_{neb})\sim RV([N\,${\sc ii}]$\lambda6584)$.  We then calculate the
mean value of $\rm RV_{sep}$ from its histogram in each plate, and pick out ten
plates with mean $\rm RV_{sep}$ greater than $\rm 25\,km\,^{-1}$ to do further
analyses. To be better describing this sample in the following text, we sort
these plates according to solar altitude (sunalt) which refers to the angle of
the sun relative to the Earth’s horizon, and assign ID number to them from 1 to
10. The choice of the $\rm RV_{sep}$ cut can be explained as follows. As the
mean value of line dispersion of $\lambda$6554 is about 0.32\,\AA\, (see
Figure~\ref{fig:skyline_sigma}), the resolution power at this wavelength is
about 8700 or resulting resolution of 34.5 $\rm km^{-1}$. However, we found
that if we constrain the parameters as more as possible in the fitting, two
components with $\rm RV_{sep}\sim 25\,km^{-1}$ still can be resolved. On the
other hand, there are only several plates having $\rm RV_{sep}>34.5\,km^{-1}$,
the lower cut can enlarge the analysing sample. 

\begin{figure*}
\centering
\includegraphics[width=\textwidth]{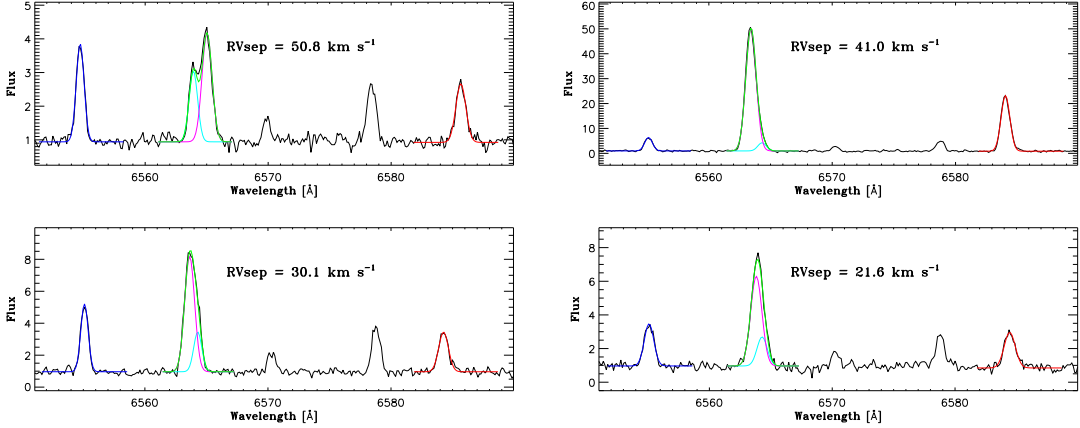}
\caption{Four examples of fitting the H$\alpha$ emission line. Both sky $\lambda$6554 (blue) and nebular [N\,{\sc ii}]$\lambda6584$ (red) are fitted using a single Gaussian. The line centroid of $\rm H\alpha_{sky}$ is fixed to have same RV with $\lambda$6554, and the line dispersion is fixed to $\sqrt{\sigma(\lambda6554)^2 + 0.02^2}$\,\AA. The line centroid of $\rm H\alpha_{neb}$ is fixed to have same RV with nebular [N\,{\sc ii}]$\lambda$6584. The best fitting result (green) which is composed of the $\rm H\alpha_{sky}$ (cyan) and the $\rm H\alpha_{neb}$ (magenta) components is obtained by minimizing the $\chi^2$. The value of RV separation between the two components is indicated in each panel.}
\label{fig:double_gaussian}
\end{figure*}

\begin{figure*}
\centering
\includegraphics[width=\textwidth]{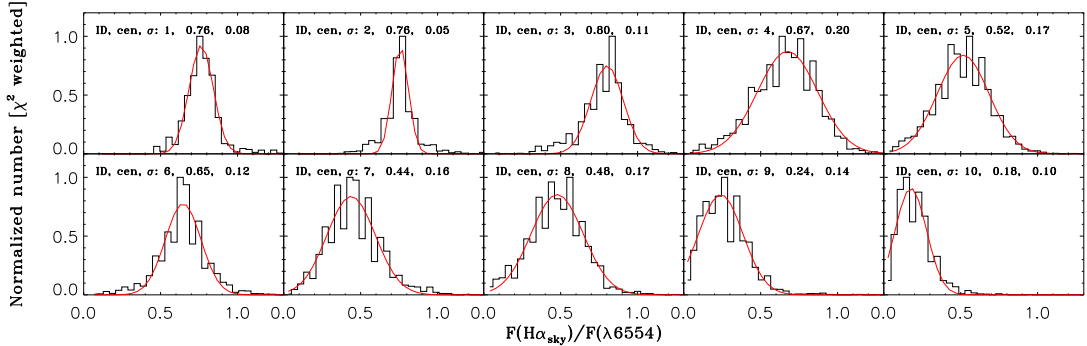}
\caption{The histograms of F(H$\rm \alpha_{sky}$)/F($\rm \lambda$6554) ratio in ten plates. The mean value and 1$\sigma$ standard deviation of the ratio are indicated in each panel.}
\label{fig:hist_ratio}
\end{figure*}

\begin{figure*}
\centering
\includegraphics[width=\textwidth]{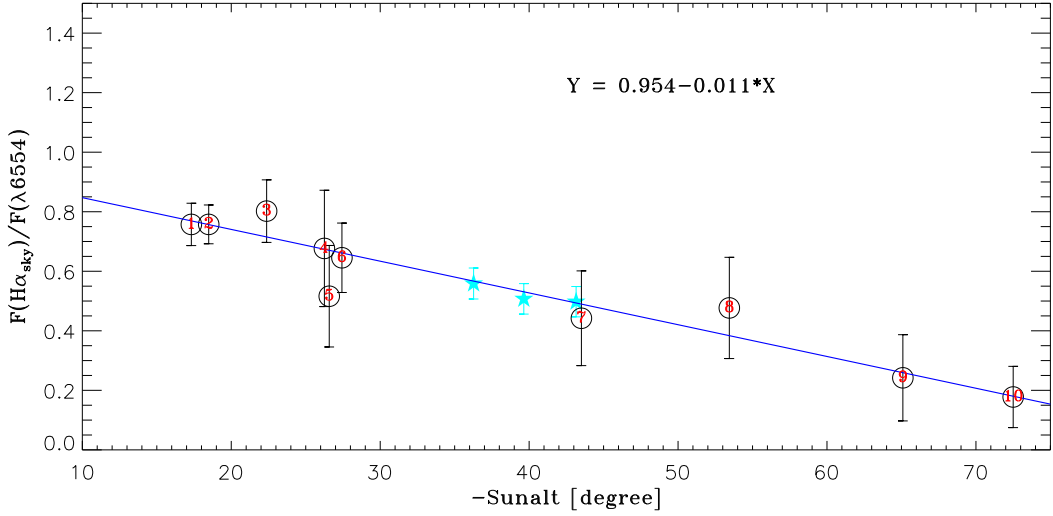}
\caption{The $\rm F(H\rm \alpha_{sky})/F(\lambda6554)$ ratio as a function of solar altitude. The ten plates are shown as open black circles with the ID number in the center, and have been used to fit the linear function. The three single exposures at the dark night are shown as cyan stars to make a comparison with the linear relation. The corresponding coefficients of this relation are indicated in the figure.}
\label{fig:ratio_sda}
\end{figure*}

We show four examples of fitting the observational $\rm H\alpha$ emission lines
in Figure~\ref{fig:double_gaussian}.  For each science spectrum, we first fit
sky line $\lambda$6554 and nebular line [N\,{\sc ii}]$\lambda$6584 using a
single Gaussian, then we use the results to constrain the fitting procedure of
the $\rm H\alpha$ emission line.  For the H$\rm \alpha_{sky}$ component, the
centroid is fixed assuming $\rm RV(H\alpha_{sky}) = RV(\lambda6554)$, the line
dispersion is fixed to $\sqrt{\sigma(\lambda6554)^2 + 0.02^2}$\,\AA, while the
line intensity is set as a free parameter.  For H$\rm \alpha_{neb}$, the
centroid is fixed to $\rm RV(H\alpha_{neb})=RV($[N\,{\sc ii}]$\lambda6584)$,
while the line dispersion and line intensity are set as free parameters.  The
best fit is obtained by minimizing the $\chi^2$, which is the goodness of the
fit of the $\rm H\alpha$ emission line. 

For each plate, we obtain the mean value of $\rm
F(H\alpha_{sky})/F(\lambda6554)$ ratio from the $\chi^2$-weighted histogram
and corresponding 1$\sigma$ standard deviation shown in Figure~\ref{fig:hist_ratio}.
We can see that $\rm F(H\rm \alpha_{sky})/F(\lambda6554)$ ratios vary from
plate to plate, spanning from 0.18 to 0.83, and the 1$\sigma$ standard
deviations are smaller than 0.2. This result confirms that this ratio cannot
be fixed for all the plates, as we described in Section 2.1 using the dark
night spectra.

\subsection{Correlation between F(H$\rm \alpha_{sky}$)/F($\rm \lambda$6554) Ratio and Solar Altitude}

Based on the observational data from the WHAM survey, the intensities of H$\rm
\alpha_{sky}$ have been found to gradually decline with the line-of-sight
shadow height \citep{Nossal-2001}. As geocoronal H$\rm \alpha_{sky}$ emission
mainly originates in the thermosphere and exosphere \citep{Mierkiewicz-2006},
while the OH $\lambda$6554 comes from the mesopause region \citep{Gardner-2017,
Xu-2012}, it's expected that F(H$\rm \alpha_{sky}$)/F($\rm \lambda$6554) ratio
should be correlated with solar altitude (sunalt). Because all the observations
were made at night, the sunalt varies from -90 to 0 degree. We show the results
derived from the ten plates in Figure~\ref{fig:ratio_sda} and find that F(H$\rm
\alpha_{sky}$)/F($\rm \lambda$6554) ratio decreases with decreasing sunalt. The
relation can be well fitted by a two-order polynomial function shown as blue
line. The coefficients of this fitting result are indicated in the figure. We
note that the three data points from the night sky field shown as filled stars
are not used to the fitting. It is found that seven of ten plates follow
the relation well, while the rest three plates (ID: 3, 5 and 8) slightly
deviate from the relation. For the two plates (ID: 5 and 8, $\rm RV_{sep}$:
25.1 and 29.5 $\rm km\,s^{-1}$), the low $\rm RV_{sep}$ might be a possible
reason of the deviation. While for the plate (ID: 3, $\rm RV_{sep}: 42.8
km\,s^{-1}$), the deviation might be caused by unusually activity of the sun,
which may change the excitation state of atmosphere at high altitude
\citep{Kerr-2001}. We note that the data points from the night sky field
follow the linear relation well, implying that although the WIM and/or the
faint nebular line can affect the line dispersion of the H$\rm \alpha$ sky
line, but the line strength of the contamination is very low compared to the
sky line.

\begin{figure*}
\centering
\includegraphics[width=\textwidth]{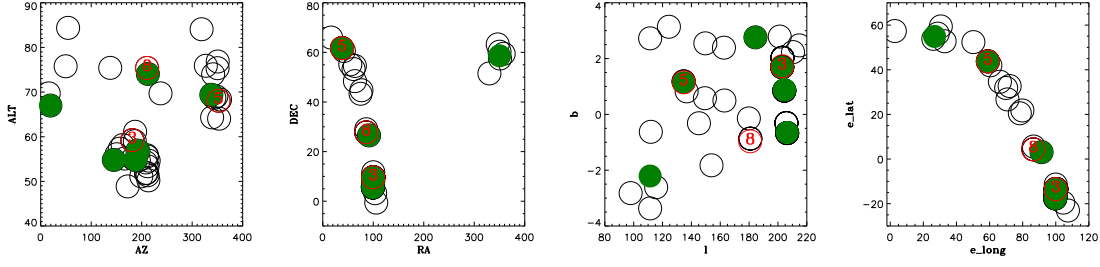}
\caption{Parameter distributions of the observed plates. The black open circles are 45 plates we have obtained, the green filled circles are the seven plates that well follow the relation between F(H$\rm \alpha_{sky}$)/F($\rm \lambda$6554) ratio and solar altitude, while the red open circles with ID numbers are the rest three plates that slightly deviate from the relation.}
\label{fig:check_sample_bias}
\end{figure*}

One may ask a question that how the ten plates can be representative of the
whole dataset, or is there any bias to apply this relation to other plates. We
here compare the parameter distributions of these ten plates to that of all 45
plates we currently have. In Figure~\ref{fig:check_sample_bias} we show  45
plates as black circles in the diagrams of right ascension versus declination,
azimuth versus altitude, Galactic longitude versus Galactic latitude, and
ecliptic longitude versus ecliptic latitude. The seven plates which follow the
relation well are shown as green filled circles, while the three plates which
are slightly deviate from the relation are shown as red open circles with ID
number in the center.  It is found that the ten plates do not prefer special
pointings. Furthermore, for each of the three ``bad" plates, it has a ``good"
plate with a comparable parameter, except the plate ID 8 in the Galactic
longitude versus Galactic latitude diagram. We then conclude that the ten
plates used to derive the relation between F(H$\rm \alpha_{sky}$)/F($\rm
\lambda$6554) ratio and sunalt can represent the whole dataset. 

\subsection{Subtracting H$\rm \alpha_{sky}$ from the Science Spectrum}

For each science spectrum, we first fit $\lambda$6554 with a single Gaussian,
and obtain the line centroid c($\lambda6554$), line dispersion
$\sigma(\lambda6554)$ and integrated flux F($\lambda$6554). Then we construct
the H$\rm \alpha_{sky}$ component using three parameters, $\rm c(\lambda6554) =
6563/6554 \times RV(\lambda6554)$, $\rm
\sigma(H\alpha_{sky})=\sqrt{\sigma(\lambda6554)^2 + 0.02^2}$\,\AA, and $\rm
F(H\alpha_{sky}) = ratio \times F(\lambda6554)$, where the ratio is the mean
value in the field, which is derived from the relation shown in Figure
\ref{fig:ratio_sda}. This sky component has been removed before we fit the rest
spectrum using a single Gaussian. We note that during the fitting, we did not
constrain the procedure anymore, instead, all the parameters are set free. In
result, we obtain line centroid, line dispersion, and line intensity for the
H$\alpha$ nebular component.

\begin{figure}
\centering
\includegraphics[width=\textwidth]{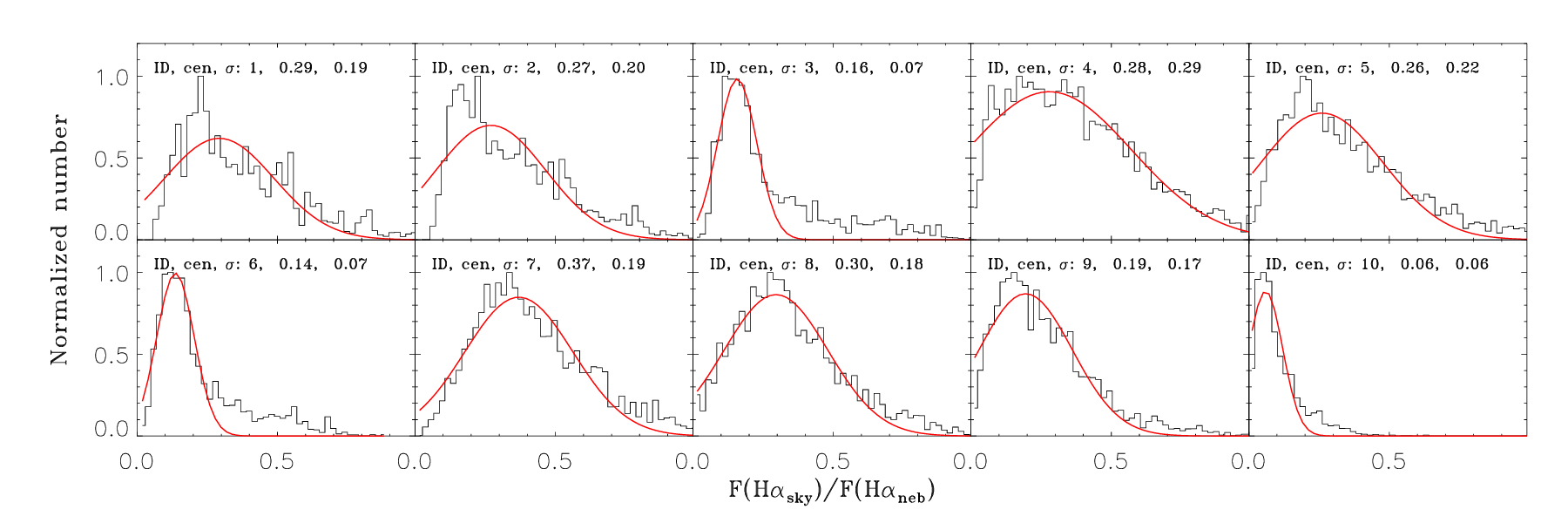}
\includegraphics[width=\textwidth]{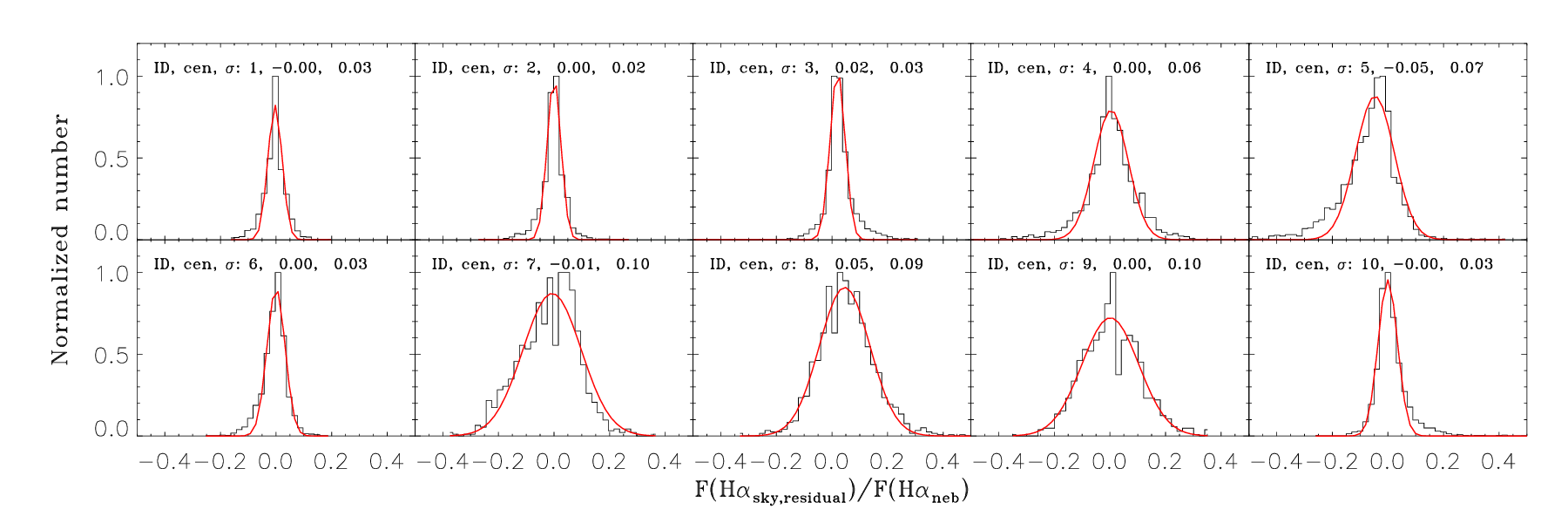}
\caption{Upper two rows: histograms of the H$\rm \alpha_{sky}$/H$\rm \alpha_{neb}$ ratios. Bottom two rows: histograms of the ratios of H$\rm \alpha$ residual to H$\rm \alpha_{neb}$.}
\label{fig:hist_sky_to_neb}
\end{figure}

We then check the residual of the sky component H$\rm \alpha_{sky,residual}$ in
Figure \ref{fig:hist_sky_to_neb}. In the upper two rows, we show the histogram
of H$\rm \alpha_{sky}$/H$\rm \alpha_{neb}$ flux ratio. It can be seen, there is
only one plate whose ratio is less than 10\%, while for the rest plates
the ratios can reach to about 40\% or even higher.  This result indicates
that the sky component is necessary to be removed, otherwise the intensity of
H$\alpha$ will be overestimated, and will seriously affect our analysis of the
physical conditions of the ISM.  In the lower two rows, we show the histograms
of H$\rm \alpha_{sky,residual}$/H$\rm \alpha_{neb}$ flux ratio. It is found
that, the ratios decrease significantly, and in most cases, the median value is
around 0., while for the three plates (ID: 3, 5 and 8) there are a little
diversion from 0 (still less than 7\%), because the mean ratios derived
from the relation are differ from the true situation. This result make us
believe that effect of H$\rm \alpha_{sky,residual}$ to the H$\rm \alpha_{neb}$
component is less than about 10\%. 

We should note that, for these ten plates, we can subtract  H$\rm \alpha_{sky}$
using the ``true" ratio (see Figure~\ref{fig:hist_ratio}). While for the other
fields that RV separation is too small that the lines cannot be resolved, we
will derive the ratio from the relation and remove the H$\rm \alpha_{sky}$
component. In most of cases, the residual flux is less than than 5\% of the
flux of H$\rm \alpha_{neb}$. Furthermore, even for the spectrum in which the
sky and nebular components can be well separated, we did not try to subtract
the sky component using double Gaussian fitting. Instead, we use a single value
of $\rm F(H\rm \alpha_{sky})/F(\lambda6554)$  ratio for the whole plate and
subtract the sky component for each fiber according to the method described
above. 

\section{Conclusion and Discussion}

Based on ten plates which the sky and nebular components can be well
resolved, we investigate the correlation between H$\rm
\alpha_{sky}$/$\lambda$6554 ratio and solar altitude, and develop a new method
to subtract geocoronal H$\alpha$ emission from the science spectrum. The main
results are summarized as follows.

\begin{itemize}

\item[-] The line dispersion of OH $\lambda$6554 is mainly caused by instrumental
broadening, while H$\rm \alpha_{sky}$ needs an extra broadening of 0.02\,\AA,
because this line is composed of two single lines with different line
centroids. 

\item[-] The blended H$\alpha$ sky and nebular lines can be resolved by two Gaussians when
the RV separation is larger than 25 $\rm km\,s^{-1}$. For this kind of plates,
F(H$\rm \alpha_{sky}$)/F($\lambda$6554) ratio can be derived directly. While it
is not true for the plates with RV separation smaller than 25 $\rm km\,s^{-1}$,
because of the limitation of the resolution power of LAMOST-MRS-N.

\item[-] The flux ratios of the H$\alpha$ sky line to the adjacent OH
$\lambda$6554 single line do not show a pattern or gradient distribution in a
plate. However, we found that the ratio decreases with decreasing solar
altitude, and this relation can be described by a linear function
and can be used to construct the H$\rm \alpha_{sky}$ spectrum for each fiber.

\item[-] The intensity of the H$\rm \alpha_{sky}$ component seriously affect the
measurements of the H$\rm \alpha_{neb}$ component. After the sky has been
subtracted, the ratio of the residual of H$\rm \alpha_{sky}$ to the
H$\rm \alpha_{neb}$ can be significantly down to about 10\%.

\end{itemize}

There are several advantages of using the relation of geocoronal H$\alpha$ to
OH $\lambda6554$ to subtract the sky component from the science spectrum: (1)
$\rm H\alpha_{sky}$ and $\lambda$6554 have very close wavelengths, hence the
uncertainties caused by wavelength calibration and responding curve can be
omitted. (2) OH $\lambda$6554 is a single line and is not polluted by
other sky or nebular lines. (3) The line intensity of $\lambda$6554 is
comparable to that of $\rm H\alpha_{sky}$. (4) Although the ratio of this two
lines changes with sunalt, but the scatter of the ratio at a specific time for
a plate is small.

This method is very covenient to be applied to the LAMOST-MRS-N data. For
normal method, there is large uncertainties in fitting the H$\alpha$
nebular+sky line using two Gaussians when these two components have similar
centroids of wavelength. Besides, it is difficult to separate the H$\alpha$ sky
component from the stars with strong H$\alpha$ absorption lines, or Be,
Wolf-Rayet stars with strong emission lines \citep{Zhang-2020,Wu-2020a}, or
SNRs with multi-component emissions \citep{Ren-2018}.  The method described in
this paper is a good way to construct the H$\alpha$ sky component from the
unblended sky line OH $\lambda6554$ and subtract it from the original spectrum.
The new generated spectrum is helpful to improve the accuracy of the 
parameters and the classifications of Galactic nebulae. Considering some
plates deviate from the F(H$\rm \alpha_{sky}$)/F($\lambda$6554) -- sunalt
relation, we admit this method has some shortcomings. A larger sample in future
is expected to find the reason of the deviation from the relation and solve
this problem.

In principle, this method can be used to subtract the sky line O{\sc i}
6300\,\AA.  However, as this line is too strong and there are no other
comparable strong single sky line nearby, the uncertainty is expected to be
higher than that of H$\alpha$. We will try to study this in another work.

\normalem
\begin{acknowledgements}

The authors thank the anonymous referee for helpful comments that improved this
manuscript. This work is supported by the National Natural Science Foundation
of China (NSFC) (no. 12090041, 12090044, 12090040, 12073051, 11733006,
11903048, 11973060) and the National Key R\&D Program of China grant (no.
2017YFA0402704). C.-H. Hsia acknowledges the support from the Science and
Technology Development Fund, MacauSAR (no.  0007/2019/A).  This work is also
supported by Key Research Program of Frontier Sciences, CAS (no.
QYZDY-SSW-SLH007) and the Guangxi Natural Science Foundation (No. 
2019GXNSFFA245008). The Guoshoujing Telescope (the Large Sky Area Multi-Object
Fiber Spectroscopic Telescope LAMOST) is a National Major Scientific Project
built by the Chinese Academy of Sciences. Funding for the project has been
provided by the National Development and Reform Commission.  LAMOST is operated
and managed by the National Astronomical Observatories, Chinese Academy of
Sciences. 

\end{acknowledgements}

\bibliographystyle{raa}
\bibliography{ref}

\label{lastpage}
\end{document}